\documentclass[unsortedaddress,superscriptaddress]{revtex4}
\usepackage{amsmath}  
\usepackage{amsbsy,amssymb,amsfonts}
\usepackage{graphicx}
\usepackage{color}
\usepackage{epsf}
\usepackage[utf8]{inputenc}
\usepackage{hyperref}
\usepackage{multirow}
\def\ie{{\em{i.e.}},}

\def\bravert{\egroup\,\vrule\,\bgroup}

{\catcode`\|=\active
  \gdef\Twoint#1{\left(\mathcode`\|"8000\let|\bravert {#1}\right)}}
{\catcode`\|=\active
  \gdef\Braket#1{\left<\mathcode`\|"8000\let|\bravert {#1}\right>}}

\newcommand{\beq}{\begin{equation}}
\newcommand{\eeq}{\end{equation}}
\newcommand{\beqa}{\begin{eqnarray}}
\newcommand{\eeqa}{\end{eqnarray}}
\newcommand{\bea}{\begin{array}}
\newcommand{\eea}{\end{array}}

\newcommand{\bef}{\begin{figure}}
\newcommand{\ef}{\end{figure}}
\newcommand{\bc}{\begin{center}}
\newcommand{\ec}{\end{center}}
\newcommand{\bt}{\begin{table}}
\newcommand{\et}{\end{table}}
\newcommand{\btb}{\begin{tabular}}
\newcommand{\etb}{\end{tabular}}

\def\etal{{\it et al.\ }}
\def\au{{\it a.u.\ }}

\def\molcas{$\cal M\kern-0.10em O\kern-0.15em L\kern-0.00em 
             C\kern-0.10em A\kern-0.05em S$}
\begin{document}

\vspace{2cm}
\title {{
         ${\cal{P,T}}$-Violating and Magnetic Hyperfine Interactions in Atomic Thallium
       }}

\vspace*{2cm}

\author{Timo Fleig}
\email{timo.fleig@irsamc.ups-tlse.fr}
\affiliation{Laboratoire de Chimie et Physique Quantiques,
             IRSAMC, Universit{\'e} Paul Sabatier Toulouse III,
             118 Route de Narbonne, 
             F-31062 Toulouse, France }
             
\author{Leonid V.\ Skripnikov}
\email{leonidos239@gmail.com}
\affiliation{National Research Centre ``Kurchatov Institute'' 
             B.P. Konstantinov Petersburg Nuclear Physics Institute, Gatchina, Leningrad District 188300, Russia}
\affiliation{Saint Petersburg State University, 7/9 Universitetskaya nab., St. Petersburg, 199034 Russia}

\date{\today}
\vspace*{1cm}
\begin{abstract}
 We present state-of-the-art configuration interaction and coupled cluster calculations of the 
 electron electric dipole moment, the nucleon-electron scalar-pseudoscalar, and the magnetic hyperfine
 interaction constants ($\alpha_{d_e}, \alpha_{C_S}, A_{||}$, respectively) for the thallium atomic ground state $^2P_{1/2}$.
 Our present best values are $\alpha_{d_e} = -559 \pm 28$, $\alpha_{C_S} = 6.77 \pm 0.34$ $[10^{-18} e$ cm$]$, and 
 $A_{||} = 21172 \pm 1059$ [MHz]. These findings lead to a significant reduction of the theoretical uncertainties for
 ${\cal{P,T}}$-odd interaction constants but not to stronger constraints on the electron electric dipole
 moment, $d_e$, or the  nucleon-electron scalar-pseudoscalar coupling constant, $C_S$.
\end{abstract}

\maketitle

\section{Introduction}
\label{SEC:INTRO}

Electric dipole moments (EDM) of elementary particles, atoms and molecules give rise to spatial parity 
(${\cal{P}}$) and time-reversal (${\cal{T}}$) violating interactions \cite{khriplovich_lamoreaux} and 
are a powerful probe for physics beyond the standard model (BSM) \cite{ramsey-musolf_review2_2013}.
Current single-source limits \cite{ACME_ThO_eEDM_nature2018,Skripnikov_ThO_JCP2016,Denis-Fleig_ThO_JCP2016} 
on the electron EDM, for instance, can probe New Physics (NP) up to an effective energy scale 
of $1000$ TeV \cite{Reece_EDMconstraint_2019} (radiative stability approach) or even greater \cite{DekensJung_JHEP2019},
surpassing the current sensitivity of the Large Hadron Collider for corresponding sources of NP.

Until today no low-energy EDM experiment has delivered a positive result. However, the obtained EDM 
upper bounds
are useful for constraining ${\cal{CP}}$-violating parameters \cite{Chupp_Ramsey_Global2015} of BSM models, 
cast as effective field theories \cite{Reece_EDMconstraint_2019,ramsey-musolf_review1_2013} at
different energy scales. 

Open-shell atomic and molecular systems are particularly sensitive probes of leptonic and semi-leptonic
  ${\cal{CP}}$-violation \cite{EDMsNP_PospelovRitz2005}.

\vspace*{0.5cm}
\noindent
\begin{minipage}{16.0cm}
 \begin{minipage}{7.0cm}
  \begin{center}
    \includegraphics[angle=0,width=6.1cm]{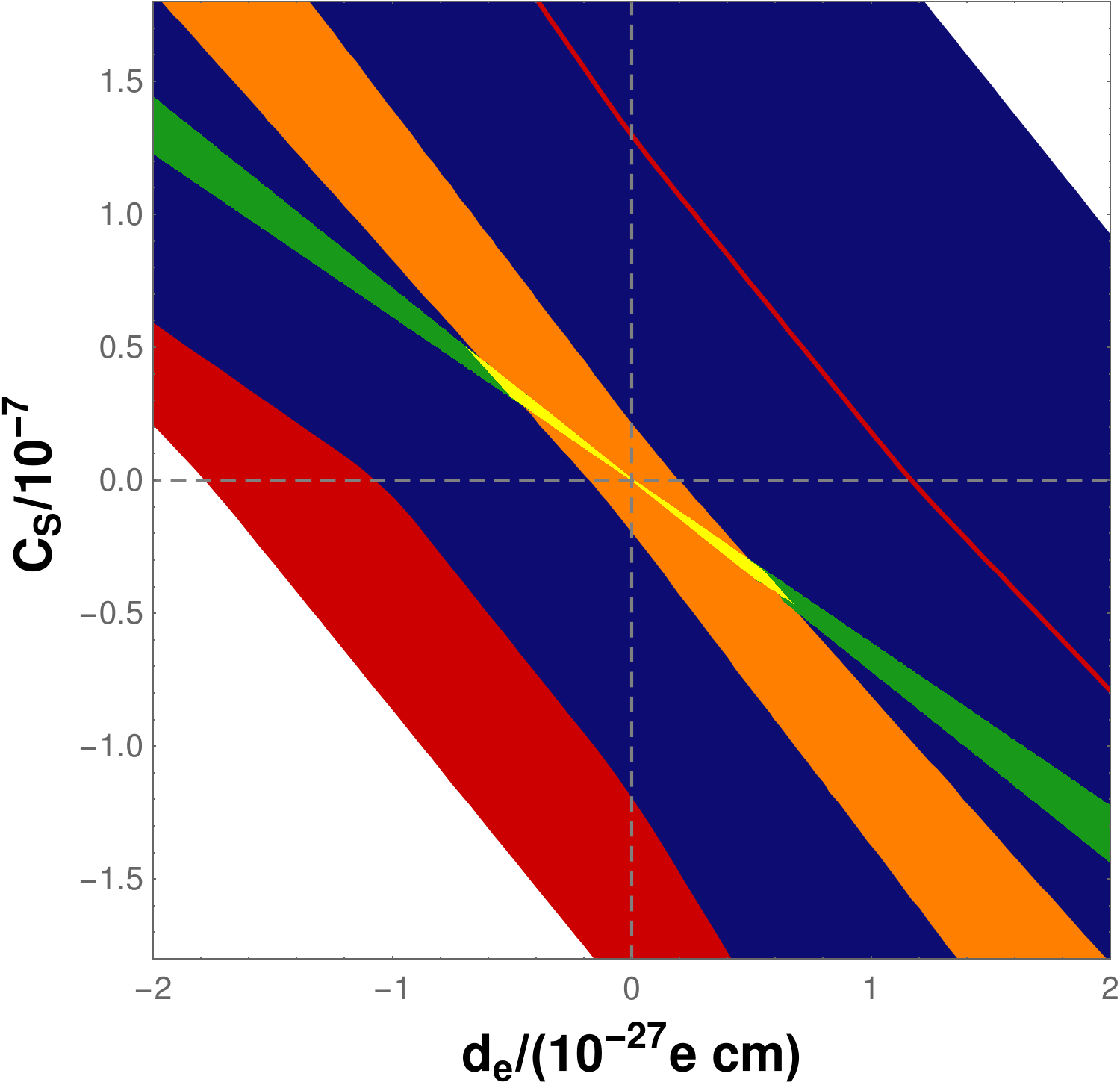}
  \end{center}
 \end{minipage}
  \hfill
 \begin{minipage}{8.8cm}
  In most BSM models \cite{Barr_eN-EDM_Atoms_1992} the dominant ${\cal{CP}}$-odd sources in open-shell systems
  are the electron EDM, $d_e$, and the nucleon-electron
  scalar-pseudoscalar (Ne-SPS) coupling, $C_S$. The panel \footnote{Courtesy: Martin Jung, Torino, Italy (2019)} shows
  the constraints (yellow surface) on $d_e$ and $C_S$ using the combined information from measurements 
  \cite{ACME_ThO_eEDM_nature2018,hinds_NJP_YbF2013,HfF+_EDM_PRL2017,regan_demille_2002}
  and calculations \cite{Skripnikov_ThO_JCP2016,Denis-Fleig_ThO_JCP2016,Sunaga_YbF_PRA2016,Abe_YbF_PRA2014,Skripnikov_HfF+_JCP2017,PhysRevA.96.040502,mar-pendrill-lindroth_EPL1991,Liu_Kelly_Tl_enhancement,dzuba_flambaum_Cs_Tl_2009,Nat_PRL_EDM_Tl,porsev_safronova_kozlov_tl2012},
  including the associated experimental and theoretical uncertainties, on the open-shell systems ThO (green), YbF (red), 
  HfF$^+$ (orange) and Tl (blue) through a global fit in the $d_e$ / $C_S$ plane. Results from a single system do, therefore,
  not constrain $d_e$ or $C_S$ individually at all in this multiple-source interpretation \cite{MJung_robustlimit2013}, 
  but lead to a fan-shaped surface of allowed combinations. The width of this surface is a function of the experimental
  and theoretical uncertainties.
  
 \end{minipage}
\end{minipage}

\vspace*{0.5cm}
This means that a substantial reduction of an uncertainty for an individual system could lead to more stringent
constraints on the unknown ${\cal{CP}}$-violating parameters. The main reason for this is that the surfaces
for different systems are not fully aligned, which is due to the different dependency of electron EDM and Ne-SPS
atomic interactions on the electric charge of the respective heavy nuclei \cite{PhysRevA.84.052108,PhysRevA.85.029901}.

A substantial part of the width of the surface for the Tl atom is due to the great spread of theoretical values
for the electron EDM atomic enhancement, $R$, calculated in the past by various groups using different 
electronic-structure approaches \cite{khriplovich_lamoreaux,johnson_tlEDM_1986,mar-pendrill-lindroth_EPL1991,Liu_Kelly_Tl_enhancement,dzuba_flambaum_Cs_Tl_2009,Nat_PRL_EDM_Tl,porsev_safronova_kozlov_tl2012}. Strikingly, Nataraj \etal\ \cite{Nat_PRL_EDM_Tl} used
a high-level many-body approach, the Coupled Cluster (CC) method, and produced a value for $R$ that strongly disagrees
with the results from all other groups, on the order of $20$\%.

The purpose of this paper is twofold:
\begin{enumerate}
 \item We use state-of-the-art Configuration Interaction (CI) and Coupled-Cluster approaches for large-scale
       applications to determine the mentioned atomic interaction constants. We put particular emphasis on the electron 
       EDM enhancement $R$ and a conclusive resolution of the major discrepancy between literature values. 
       Claims about physical effects that purportedly underlie these discrepancies are scutinized.
 \item We investigate whether a reduced uncertainty for $R$(Tl) impacts the above-described constraints on
       $d_e$ and $C_S$.
\end{enumerate}
The paper is structured as follows. In section \ref{SEC:THEORY} we lay out the theory underlying the atomic electron EDM,
Ne-SPS, and magnetic hyperfine interactions constants. The following section \ref{SEC:RESULTS} contains technical details
about our calculations, results, and a discussion of these results in comparison with literature values.
The final section \ref{SEC:CONCL} concludes on our findings.

\section{Theory}
\label{SEC:THEORY}

An atomic electric dipole moment (EDM) is defined \cite{commins_EDM_1999} (p. 16) as 
\begin{equation}
 d_a = -\lim\limits_{E_{\text{ext}} \rightarrow 0}\, \left[\frac{\partial (\Delta \varepsilon)}{\partial E_{\text{ext}}} \right]
 \label{EQ:ATOMIC_EDM}
\end{equation}
where $\Delta \varepsilon$ is a ${\cal{P,T}}$-odd energy shift and $E_{\text{ext}}$ is an external electric field. 
In atoms with nuclear spin $I \leq \frac{1}{2}$ \cite{Sushkov_Flambaum_Khriplovich1984}
and in an electronic state with unpaired electrons, this energy shift is dominated by and originates from either the electron EDM, $d_e$, or a ${\cal{P,T}}$-odd nucleon-electron (Ne) interaction, or a combination of the two \cite{Barr_eN-EDM_Atoms_1992,EDMsNP_PospelovRitz2005}. 
The two cases are presented separately.

\subsection{Atomic EDM due to electron EDM}

The Hamiltonian for the interaction of the electron electric dipole moment, $d_e$, is for an atomic system
\begin{equation}
 H_{\text{EDM}} = -\sum\limits_j\, {\bf{d}}_j \cdot {\bf{E}}({\bf{r}}_j)
   = -d_e\, \sum\limits_j\, \gamma^0_j\, \boldsymbol{\Sigma}_j\cdot{\bf{E}}({\bf{r}}_j)
\end{equation}
where $\gamma^0$ is a Dirac matrix, $\boldsymbol{\Sigma} = \left( 
\begin{array}{cc} \boldsymbol{\sigma} & {\bf{0}} \\ {\bf{0}} & \boldsymbol{\sigma} \end{array} \right)$ is a vector
of spin matrices in Dirac representation, $j$ is an electron index, ${\bf{E}}({\bf{r}}_j)$ the electric field at position 
${\bf{r}}_j$ and the bare fermion's electric dipole moment is expressed as 
${\bf{d}} = d_e\, \gamma^0\, \boldsymbol{\Sigma}$, necessarily linearly dependent on the particle's spin vector 
$\boldsymbol{\Sigma}$ \cite{khriplovich_lamoreaux,Hunter_Science}.

Supposing a non-zero electron EDM $d_e$, the resulting energy shift can be evaluated as
\begin{equation}
 \Delta \varepsilon_{\text{EDM}} = 
 d_e\, \left< -\sum\limits_j\, \gamma^0_j\, \boldsymbol{\Sigma}_j\cdot{\bf{E}}({\bf{r}}_j) \right>_{\psi(E_{\text{ext}})}
 \label{EQ:ENSHI_EXPEC}
\end{equation}
where $\psi(E_{\text{ext}})$ is the field-dependent atomic wavefunction of the state in question. 
The expectation value in Eq. (\ref{EQ:ENSHI_EXPEC})
has the physical dimension of electric field and can be regarded as the mean interaction of each electron EDM with
this field in the respective state. Following stratagem II of Lindroth {\etal} \cite{lindroth_EDMtheory1989} the
expectation value is recast in electronic momentum form as an effective one-body operator
\begin{equation}
\left< -\sum\limits_j\, \gamma^0_j\, \boldsymbol{\Sigma}_j\cdot{\bf{E}}({\bf{r}}_j) \right>_{\psi(E_{\text{ext}})}
\approx \frac{2\imath c}{e\hbar}\, \left< \sum\limits_j\, \gamma^0_j\, \gamma^5_j\, {\bf{p}}_j^2 \right>_{\psi(E_{\text{ext}})}
 \label{EQ:STRATAGEMII}
\end{equation}
where the approximation lies in assuming that $\psi$ is an exact eigenfunction of the field-dependent Hamiltonian
of the system. This momentum-form EDM operator has already been used as early as in 1986, by Johnson \etal\
\cite{johnson_tlEDM_1986}.
In the present work the field-dependent Hamiltonian is the Dirac-Coulomb (DC) Hamiltonian (in 
\au, $e = m_e = \hbar = 1$)
\begin{eqnarray}
 \nonumber
 \hat{H} &:=& \hat{H}^{\text{Dirac-Coulomb}} + \hat{H}^{\text{Int-Dipole}} \\
  &=& \sum\limits^n_j\, \left[ c\, \boldsymbol{\alpha}_j \cdot {\bf{p}}_j + \beta_j c^2 
- 
  \frac{Z}{r_{jK}}{1\!\!1}_4 \right]
+ \sum\limits^n_{k>j}\,
\frac{1}{r_{jk}}{1\!\!1}_4
  + \sum\limits_j\, {\bf{r}}_j \cdot {\bf{E}_{\text{ext}}}\, {1\!\!1}_4
 \label{EQ:HAMILTONIAN}
\end{eqnarray}
with ${\bf{E}_{\text{ext}}}$ weak and homogeneous, the indices $j,k$ run over $n$ electrons, $Z$ the proton 
number with the nucleus $K$ placed at the origin, and $\boldsymbol{\alpha}$ are standard Dirac matrices.
$E_{\text{ext}}$ is not treated as a perturbation but included {\it{a priori}} in the variational optimization of
the atomic wavefunction. Furthermore, the final results reported in this work include high excitation ranks in the 
correlation expansion of $\psi$. For these reasons, the approximation in Eq. (\ref{EQ:STRATAGEMII}) is considered
very good in the present case.

Within the so-defined picture and using Eqs. (\ref{EQ:ATOMIC_EDM}), (\ref{EQ:ENSHI_EXPEC}), and 
(\ref{EQ:STRATAGEMII}) the atomic EDM becomes
\begin{equation}
 \label{EQ:ATOMIC_EDM_LII}
 d_a = -\lim\limits_{E_{\text{ext}} \rightarrow 0}\,
       \frac{\partial}{\partial E_{\text{ext}}}
       \frac{2\imath c\, d_e}{e\hbar}\, \left< \sum\limits_j\, \gamma^0_j\, \gamma^5_j\, 
            {\bf{p}}_j^2 \right>_{\psi(E_{\text{ext}})}.
\end{equation}
The (dimensionless) atomic EDM enhancement factor is defined as $R := \frac{d_a}{d_e}$. Denoting
$E_{\text{eff}} = \frac{2\imath c}{e\hbar}\, \left< \sum\limits_j\, \gamma^0_j\, \gamma^5_j\, {\bf{p}}_j^2 \right>_{\psi(E_{\text{ext}})}$
for the sake of simplicity, the enhancement factor is
\begin{equation}
 \label{EQ:R_IN_EFF}
 R = -\lim\limits_{E_{\text{ext}} \rightarrow 0}\, \left[ \frac{\partial E_{\text{eff}}}{\partial E_{\text{ext}}} \right].
\end{equation}
The external field used in the experiment on Tl \cite{regan_demille_2002} was $E_{\text{ext}} = 1.23 \times 10^7 
\left[ \frac{\text{V}}{\text{m}} \right] \approx 0.2392 \times 10^{-4}$ a.u. 
In the present work $E_{\text{ext}} = 0.24 \times 10^{-4}$ a.u. is used. This is a very small field which is well 
within the linear regime considering the derivative in Eq. (\ref{EQ:R_IN_EFF}). The enhancement factor may under these
circumstances be written as a function of two field points
\begin{equation}
 R_{\text{lin}} = -\frac{\Delta E_{\text{eff}}}{\Delta E_{\text{ext}}} = -\frac{E_{\text{eff}}(2) - E_{\text{eff}}(1)}
                   {E_{\text{ext}}(2) - E_{\text{ext}}(1)}.
\end{equation}
We set $E_{\text{ext}}(1) := 0$ from which it follows that $E_{\text{eff}}(1) = 0$, and so
\begin{equation}
 R \approx R_{\text{lin}} = -\frac{E_{\text{eff}}}{E_{\text{ext}}}.
\end{equation}
$E_{\text{eff}}$ is calculated as described in reference \cite{fleig_nayak_eEDM2013}. ${\psi}$ is an approximate 
configuration interaction
(CI) eigenfunction of the Dirac-Coulomb Hamiltonian including $E_{\text{ext}}$.
Alternatively, $E_{\text{eff}}$ can be calculated within the finite-field approach~\cite{Skripnikov:17a,Skripnikov:11a}. The latter has been used in coupled cluster calculations.

The electron EDM enhancement factor $R$ is in the particle physics literature often denoted as
\begin{equation}
 \alpha_{d_e} := R,
\end{equation}
the atomic-scale interaction constant of the electron EDM.

\subsection{Nucleon-Electron Scalar-Pseudoscalar Interaction}

The effective Hamiltonian for a ${\cal{P,T}}$-odd nucleon-electron scalar-pseudoscalar interaction is written as
\cite{Flambaum_Khriplovich1985}
\begin{equation}
\label{SPSHam}
 H_{\text{Ne-SPS}} = \imath \frac{G_F}{\sqrt{2}}\, 
                     A C_S\, \sum\limits_j\, \gamma^0_j\, \gamma^5_j\, \rho_N({\bf{r}}_j)
\end{equation}
and the resulting atomic energy shift is accordingly
\begin{equation}
 \Delta \varepsilon_{\text{Ne-SPS}} = \frac{G_F}{\sqrt{2}}\, A C_S\,
 \left< \imath \sum\limits_j\, \gamma^0_j\, \gamma^5_j\, \rho_N({\bf{r}}_j) \right>_{\Psi(E_{\text{ext}})},
\end{equation}
where $A$ is the nucleon number, $C_S$ is the S-PS nucleon-electron coupling constant, $G_F$ is the Fermi 
constant\footnote{A comment on units:
Its value is $\frac{G_F}{(\hbar c)^3} = 1.166364 \times 10^{-5} {\text{[GeV]}}^{-2} = 0.86366 \times 10^{-20} 
E_H^{-2}$. With $\hbar = 1.$ a.u. and 
$c = 137.036$ a.u., the Fermi constant is also expressed as $G_F = 2.2225 \times 10^{-14}$ a.u.}
and $\rho_N({\bf{r}}_j)$ is the nucleon density at the position of electron $j$.
Note that in the present work we define $\gamma^5 := \imath \gamma^0 \gamma^1 \gamma^2 \gamma^3$, whereas
Flambaum and co-workers \cite{dzuba_flambaum_Cs_Tl_2009,PhysRevA.84.052108,PhysRevA.85.029901} define
$\gamma^5 := -\imath \gamma^0 \gamma^1 \gamma^2 \gamma^3$ which explains the sign difference between the
present Ne-SPS atomic interaction constants and those of Flambaum and co-workers.


Next, we define (see also reference \cite{shukla_PRA_Rb}) in analogy with Eq. (\ref{EQ:R_IN_EFF}) an Ne-SPS ratio
\footnote{The physical dimension of the $S$ ratio is dim$(S) =$ dim$\left( \frac{\rho_N}{E} \right) = \left[
\frac{Q\, T^2}{M\, L^4} \right]$. This is consistent with the dimension of $S$ in the definition, Eq. (\ref{EQ:S_DEF}),
where dim$(S) =$ dim$\left( \frac{d_a}{C_S\, G_F} \right) = \left[ \frac{Q\, T^2}{M\, L^4} \right]$.}
\begin{equation}
 S := \frac{d_a}{A C_S \frac{G_F}{\sqrt{2}}}
 \label{EQ:S_DEF}
\end{equation}
and so one can write, using Eq. (\ref{EQ:ATOMIC_EDM}),
\begin{equation}
 S = -\lim\limits_{E_{\text{ext}} \rightarrow 0}\, \left[ \frac{\partial}{\partial E_{\text{ext}}} 
 \left< \imath \sum\limits_j\, \gamma^0_j\, \gamma^5_j\, \rho_N({\bf{r}}_j) \right>_{\Psi(E_{\text{ext}})}\right]
\end{equation}
and in the linear regime
\begin{equation}
S = - \frac{\left< \imath \sum\limits_j\, \gamma^0_j\, \gamma^5_j\, \rho_N({\bf{r}}_j) \right>_{\Psi(E_{\text{ext}})}}{E_{\text{ext}}}.
\end{equation}
The initial implementation of this expectation value in the latter expression has been described in reference 
\cite{ThF+_NJP_2015}.
The independent implementation of the matrix elements of the Hamiltonian (\ref{SPSHam}) has been developed in ref. \cite{Skripnikov_ThO_JCP2016}.

For comparison with literature results we also define the S-PS nucleon-electron interaction constant
\begin{equation}
 \alpha_{C_S} := \frac{d_a}{C_S} = S\, A\, \frac{G_F}{\sqrt{2}}.
\end{equation}

\subsection{Magnetic Hyperfine Interaction}

Minimal substitution according to ${\bf{p}} \longrightarrow {\bf{p}} - \frac{q}{c}{\bf{A}}$ in the Dirac equation
and representing the vector potential in magnetic dipole approximation as 
${\bf{A}}_D({\bf{r}}) = \frac{{\bf{m}} \times {\bf{r}}}{r^3}$ with ${\bf{m}}$ the nuclear magnetic dipole moment
leads to the magnetic hyperfine Hamiltonian
\begin{equation}
 \nonumber
 \hat{H}_{\text{HF}} = c {\boldsymbol{\alpha}} \cdot \left( - \frac{q}{c}\, 
                          \frac{{\bf{m}} \times {\bf{r}}}{r^3} \right)
	             = q\, {\bf{m}} \cdot \left( \frac{{\boldsymbol{\alpha}} \times {\bf{r}}}{r^3} \right)
 \label{EQ:HAM_HF_1}
\end{equation}
for a single point charge $q$ at position ${\bf{r}}$ outside the finite nucleus. Given the nuclear magnetic dipole 
moment vector as
${\bf{m}} = \frac{\mu}{I}\, \mu_N\, {\bf{I}} = g_I\, \mu_N\, {\bf{I}}$ where $\mu$ is the magnetic moment in 
nuclear magnetons ($\mu_N$), $g_I$ is the nuclear $g$-factor, and ${\bf{I}}$ is the nuclear spin, Eq. 
(\ref{EQ:HAM_HF_1}) for a single electron is written as
\begin{equation}
 \hat{H}_{\text{HF}} = -e\, \frac{\mu}{I}\, \mu_N\, {\bf{I}} \cdot 
                       \left( \frac{{\boldsymbol{\alpha}} \times {\bf{r}}}{r^3} \right)
 \label{EQ:HAM_HF_2}
\end{equation}
Based on Eq. (\ref{EQ:HAM_HF_2}) we now define the magnetic hyperfine interaction constant for $n$ electrons in 
the field of nucleus $K$ (in \au)
\begin{equation}
 A_{||}(K) = -\frac{\mu_{K}[\mu_N]}{2cIm_p M_J}\,
 \left< \Psi_{J,M_J} \right| \sum\limits_{i=1}^n\, \left( \frac{\boldsymbol{\alpha}_i \times {\bf{r}}_{iK}}{r_{iK}^3}
 \right)_z \left| \Psi_{J,M_J} \right>
 \label{EQ:A_CONST}
\end{equation}
%
%
where $\frac{1}{2cm_p}$ is the nuclear magneton in \au\ and $m_p$ is the proton rest mass. The term $\frac{1}{M_J}$
in the prefactor of Eq. (\ref{EQ:A_CONST}) is explained as follows.

The vector operator $\left( \frac{\boldsymbol{\alpha}_i \times {\bf{r}}_{iK}}{r_{iK}^3} \right)_z$ can be regarded 
as the $q=0$ component of a rank $k=1$ irreducible tensor operator $\hat{T}^{(k)}_q$.
Application of the Wigner-Eckart Theorem to the diagonal matrix element in Eq. (\ref{EQ:A_CONST}) yields
\begin{eqnarray*}
\left< \alpha, J, M_J | \hat{T}^{(1)}_0 | \alpha, J, M_J \right> &=& \left< J, M_J; 1, 0 | J, 1; J, M_J \right>\,
                                       \frac{\left< \alpha, J || \hat{T}^{(1)} || \alpha, J \right>}{\sqrt{2J+1}}
\end{eqnarray*}
where the Clebsch-Gordan coefficient is -- using the general definition in Ref. \cite{weissbluth}, p. $27$ --
evaluated as
\begin{equation}
 \left< J, M_J; 1, 0 | J, 1; J, M_J \right> = M_J\, \frac{1}{\sqrt{J(J+1)}},
 \label{EQ:CG1}
\end{equation}
which depends linearly on the total electronic angular momentum projection quantum number $M_J$.
However, the magnetic hyperfine energy has to be independent of $M_J$ which is assured by the above prefactor
$\frac{1}{M_J}$.
%
%
%
%
%
Magnetic hyperfine interaction matrix elements have been calculated based on the implementations in references \cite{Fleig2014,Skripnikov_ThO_JCP2016} which do
 not make direct use of the Wigner-Eckart theorem and reduced matrix elements.


\section{Results and Discussion}
\label{SEC:RESULTS}

\subsection{Technical Details}

Gaussian atomic basis sets of double-, triple-, and quadruple-$\zeta$ quality 
\cite{dyall_p-basis,dyall_p,4p-basis-dyall-2} (including correlating functions for $4f$ and $5d$ shells in the
case of CI and cvDZ/CC) \cite{5fbasis-dyall-ccorr} have been used in the present work.

The atomic spinor basis is obtained in Dirac-Coulomb (DC) Hartree-Fock (HF) approximation where the Fock 
operator is defined by averaging over $6p_{j=1/2}^1$ and $6p_{j=3/2}^1$ open-shell electronic configurations.

A locally modified version of the \verb+DIRAC+ program package \cite{DIRAC16} has been used for all
electronic-structure calculations.
Interelectron correlation effects are taken into account through Configuration Interaction (CI) theory
as implemented in the \verb+KRCI+ module \cite{knecht_luciparII} of \verb+DIRAC+.
Coupled cluster (CC) calculations have been carried out within the  {\sc mrcc} code \cite{MRCC2013,Kallay:1,Kallay:2}.

The nomenclature for both CI and CC models is defined as: S, D, T, etc. denotes Singles, Doubles,
Triples etc. replacements with respect to the 
reference DCHF determinant. The following number is the 
number of correlated electrons and encodes which occupied shells are included in the CI or CC expansion.
In detail we have
$3 \mathrel{\widehat{=}} (6s,6p)$,
$13 \mathrel{\widehat{=}} (5d,6s,6p)$,
$21 \mathrel{\widehat{=}} (5s,5p,5d,6s,6p)$,
$29 \mathrel{\widehat{=}} (4s,4p,5s,5p,5d,6s,6p)$,
$31 \mathrel{\widehat{=}} (4d,5s,5p,5d,6s,6p)$,
$35 \mathrel{\widehat{=}} (4f,5s,5p,5d,6s,6p)$.
$81 \mathrel{\widehat{=}} (1s,2s,2p,3s,3p,3d,4f,5s,5p,5d,6s,6p)$.
The notation type S10\_SD13, as an example, means that the model SD13 has been approximated by
omitting Double excitations from the $(5d)$ shells. CAS3in4 means that an active space is used with all
possible determinant occupations distributing the $3$ valence electrons over the $4$ valence Kramers 
pairs.

We use the experimental value \cite{stone_INDC2015} for the nuclear magnetic moment of {$^{205}$Tl} with nuclear spin $I=\frac{1}{2}$, 
$\mu = 1.63821$[$\mu_N$], in calculations of the magnetic hyperfine interaction constant.

\subsection{Results for Atomic Interaction Constants}

The results from the systematic study of many-body effects on atomic EDM enhancement ($R$), 
Ne-SPS interaction ratio ($S$) and magnetic hyperfine interaction constant ($A$) are compiled in Table \ref{TAB2}. 
The general strategy is to first qualitatively investigate the relative importance of various 
many-body effects on the properties using a rather small atomic basis set. Then, in a second step, 
accurate models are developed that include all important many-body effects using the insight from the 
first step and larger atomic basis sets. Since EDM enhancement and Ne-SPS interaction ratio are 
analytically related \cite{PhysRevA.84.052108,PhysRevA.85.029901} it is sufficient to discuss the trends 
for $R$ only.

\begin{table}
\caption{R, S, and A for Tl atom. By default, calculations were performed using DCHF spinors for the
         neutral Tl atom ($V^N$ potential) and, for comparison in selected cases, with the Tl$^+$ cation
         ($V^{N-1}$ potential) and Tl$^{3+}$ cation ($V^{N-3}$ potential) spinors.\label{TAB2} }
  \begin{tabular}{l|ccc}
   Model/virtual cutoff                         & $R$     & $S$ [a.u.] & $A_{||}$({$^{205}$Tl}) [MHz]    \\ \hline\hline
    \multicolumn{4}{c}{Dyall cvDZ} \\
      CAS1in3                                      & $-388$  & $269$ & $18800$ \\
      CAS3in4                                      & $-415$  & $288$ & $18800$ \\
      CAS3in4\_SD3/60au                            & $-487$  & $339$ & $19092$ \\
      CAS3in4\_SDT3/60au                           & $-487$  & $339$ & $19103$ \\
      S10\_CAS3in4\_SD13/10au                      & $-458$  & $321$ & $20003$ \\
      SD10\_CAS3in4\_SD13/10au                     & $-442$  & $309$ & $19502$ \\
      SD10\_CAS3in4\_SD13/30au                     & $-441$  & $309$ & $19575$ \\
      SD10\_CAS3in4\_SDT13/10au                    & $-465$  & $326$ & $19357$ \\
      SD10\_CAS3in4\_SDTQ13/10au                   & $-464$  & $326$ & $19345$ \\
      SDT10\_CAS3in4\_SDT13/10au                   & $-460$  & $323$ & $19254$ \\
      SDT10\_CAS3in4\_SDTQ13/10au                  & $-460$  & $323$ & $19341$ \\
      SD18\_CAS3in4\_SD21/10au                     & $-437$  & $307$ & $19445$ \\
      SD18\_CAS3in4\_SD21/10au(Tl$^+$)             & $-428$  & $300$ & $18934$ \\
      S8\_SD18\_CAS3in4\_SD29/10au                 & $-438$  & $308$ & $19536$ \\
      SD18\_CAS3in4\_SD21/30au                     & $-443$  & $311$ & $19758$ \\
      SD18\_CAS3in4\_SD21/60au                     & $-443$  & $311$ & $19759$ \\
      SD8\_SD18\_CAS3in4\_SD29/30au                & $-449$  & $315$ & $19980$ \\
      SD18\_CAS3in4\_SDT21/10au                    & $-473$  & $331$ & $19439$ \\
      SD18\_CAS3in4\_SDT21/10au(Tl$^+$)            & $-467$  & $328$ & $19228$ \\
      SDT18\_CAS3in4\_SDT21/10au                   & $-461$  & $325$ & $19274$ \\
      SD18\_CAS3in4\_SDT21/30au                    & $-483$  & $338$ & $19761$ \\
      SD18\_CAS3in4\_SDT21/60au                    & $-483$  & $338$ & $19763$ \\
      S10\_SD18\_CAS3in4\_SDT31/10au               & $-469$  & $329$ & $19423$ \\
      S14\_SD18\_CAS3in4\_SDT35/10au               & $-469$  & $330$ & $19448$ \\
      S8\_SD18\_CAS3in4\_SDT29/30au                & $-484$  & $340$ & $19999$ \\
      SD8\_SDT10\_CAS3in4\_SDT21/10au              & $-471$  & $331$ &         \\
      SD18\_CAS3in4\_SDTQ21/10au                   & $-469$  & $329$ & $19395$ \\  \hline
      \multicolumn{4}{c}{Dyall cvTZ} \\
        CAS3in4                                      & $-460$  & $323$ & $     $ \\
        CAS3in4\_SD3/10au                            & $-565$  & $397$ & $19027$ \\
        CAS3in4\_SD3/50au                            & $-565$  & $397$ & $19041$ \\
        CAS3in4\_SDT3/50au                           & $-566$  & $398$ & $19050$ \\
        SD18\_CAS3in4\_SD21/10au                     & $-481$  & $340$ & $19619$ \\
        SD18\_CAS3in4\_SD21/30au                     & $-484$  & $342$ & $19751$ \\
        SD18\_CAS3in4\_SDT21/10au                    & $-542$  & $383$ & $19995$ \\
%
        SD18\_CAS3in4\_SDT21/10au(Tl$^{3+}$)         & $-524$  & $371$ &         \\
        SD18\_CAS3in4\_SDT21/20au                    & $-541$  & $383$ &         \\ \hline
        \multicolumn{4}{c}{Dyall cvQZ} \\
   CAS1in3                                      & $-429$  & $301$ & $18806$ \\
   CAS3in4                                      & $-476$  & $334$ & $18806$ \\
   CAS3in4\_SD3/10au                            & $-587$  & $412$ & $19023$ \\
   CAS3in4\_SD3/35au                            & $-587$  & $412$ & $19050$ \\
   CAS3in4\_SDT3/35au                           & $-587$  & $413$ & $19060$ \\
   SD18\_CAS3in4\_SD21/35au                     & $-459$  & $322$ & $17442$ \\
   SD18\_CAS3in4\_SDT21/10au                    & $-555$  & $391$ & $20432$ \\
   SD18\_CAS3in4\_SDT21/35au                    & $-562$  & $397$ & $20592$ \\ \hline\hline
   {\bf{cvQZ/SD18\_CAS3in4\_SDT21/35au + $\Delta_{\text{corr}}$}}   & {\bf{-539}}  & {\bf{388}} & {\bf{20614}} \\ \hline\hline
 \end{tabular}
\end{table}

\subsection{Step 1: Many-Body effects in cvDZ basis}

\subsubsection{Valence electron correlation}
The result of $R=-388$ for CAS1in3 which is a singles CI expansion for the electronic ground state can be regarded
as close to a DC Hartree-Fock result.
The Full-CI (FCI) result including only the three 
valence electrons 
(CAS3in4\_SDT3/60au) of $R=-487$ shows that valence correlation effects lead to a considerable change by 
more than $25$\% (in the large cvQZ basis by more than $35$\%). The valence FCI enhancement in cvQZ basis 
of $R=-587$ is, therefore, a benchmark. This value is closely reproduced using the universal basis set of 
reference \cite{Nat_PRL_EDM_Tl}. Further effects can be considered as modifications of this benchmark 
result and will be studied one by one.

\subsubsection{Subvalence electron correlation}
Subvalence electrons of the Tl atoms are those occuping the $5s$, $5p$, and $5d$ shells. All other 
electrons will be considered core electrons. Correlations among the $5d$ electrons and in particular of 
the $5d$ and the valence electrons lead to a strong decrease of $R$, on the absolute, on the order of 
$10$\%. Corresponding contributions from the $5s$ and $5p$ electrons are significantly smaller. 

\subsubsection{Outer-core electron correlation}
Outer-core-valence correlations have been evaluated by allowing for one hole in the respective outer
core spinors along with exctations from the subvalence and valence electrons. In sum for the shells
with effective principal quantum number $n=4$ these effects amount to about $1.5$\%.

\subsubsection{Effect of higher excitation ranks}
Allowing for three holes in the shells with effective principal quantum number $n=5$ and up to four
particles in the virtual spinors (\ie\ adding combined quadruple excitations) leads to a total change
of around $3.5$\%. 

\subsection{Step 2: Accurate CI results}
Subsets of important CI models based on the findings of the previous subsection have been repeated using 
the larger atomic basis sets, cvTZ and cvQZ. The single best values from these calculations are given
by the model SD18\_CAS3in4\_SDT21/35au. These latter values $V$ are then corrected by a ``correction 
shift'', calculated as follows:
\begin{eqnarray*}
\Delta_{\text{corr}} &:=&  V({\rm{S10\_SD18\_CAS3in4\_SDT31/10au}}) - V({\rm{SD18\_CAS3in4\_SDT21/10au}}) \\
                       &&+ V({\rm{S14\_SD18\_CAS3in4\_SDT35/10au}}) - V({\rm{SD18\_CAS3in4\_SDT21/10au}}) \\
                       &&+ V({\rm{S8\_SD18\_CAS3in4\_SDT29/30au}}) - V({\rm{SD18\_CAS3in4\_SDT21/30au}}) \\
                       &&+ V({\rm{SDT18\_CAS3in4\_SDT21/10au}}) - V({\rm{SD18\_CAS3in4\_SDT21/10au}}) \\
                       &&+ V({\rm{SD18\_CAS3in4\_SDTQ21/10au}}) - V({\rm{SD18\_CAS3in4\_SDT21/10au}})
\end{eqnarray*}
The final best CI values are obtained by adding the above sum of individual corrections to the value
from the model SD18\_CAS3in4\_SDT21/35au.

\subsection{Accurate CC results}

Table \ref{TABcc1} gives values of R, S and $A_{||}$({$^{205}$Tl}) constants obtained within the 
all-electron coupled cluster with single, double and non-iterative triple cluster amplitudes, CCSD(T), 
method employing several basis sets. One can see a good convergence of the results in the series of the 
Dyall's DZ, TZ and QZ basis sets: values of R obtained within the QZ and TZ basis sets differ by about 
2\%. Table \ref{TABcc1} also gives values of the constants obtained within the Nataraj's universal basis 
set \cite{Nat_PRL_EDM_Tl}.
Note that the latter basis set is the even-tempered basis set (geometry progression).
One can see a good agreement of the results obtained within the QZ basis set and Nataraj's universal 
basis set.

\begin{table}
 \caption{R, S, and A for Tl atom calculated within the 81e-CCSD(T) method in different basis sets. In the case denoted
	``$V^N$'' the atomic spinors are obtained for the neutral Tl atom and the external field perturbs both the
	spinor coefficients and the CC amplitudes. In the case
	denoted ``$V^{N-1}$'' the atomic spinors are obtained for the Tl$^+$ cation and the external electric field 
	only perturbs the CC amplitudes but not the atomic spinors.}
\label{TABcc1} 
\begin{tabular}{llll}
\hline
\hline
 Basis set/virtual cutoff   & $R$     & $S$ [a.u.] & $A_{||}$({$^{205}$Tl}) [MHz]    \\ \hline\hline
	Nataraj universal/$10^3$au ($V^N$)     & -559 & 397 & 21087 \\
	Nataraj universal/$10^3$au ($V^{N-1}$) & -550 & 390 & 21071 \\
\hline 
Dyall cvDZ/$10^4$au ($V^N$)& -493 & 347 & 20626  \\
Dyall cvTZ/$10^4$au ($V^N$)& -545 & 387 & 20760 \\
Dyall cvQZ/$10^4$au ($V^N$)& -558 & 397 & 21172 \\
\hline\hline
\end{tabular}
\end{table}

Table \ref{TABcc3} gives values of $R$ calculated with different number of correlated electrons. As can be seen contributions from subvalence and outer-core electrons are close to those obtained within the CI approach above.

\begin{table}
	\caption{\label{TABcc3}R for Tl atom calculated within the CCSD(T) method in Dyall's cvQZ basis set.}
 \begin{tabular}{ll}
 \hline
 \hline
  Method/virtual cutoff   & $R$   \\ 
  \hline\hline
  3e-CCSD(T)/$10$au    & -589  \\
  21e-CCSD(T)/$150$au  & -527  \\
  53e-CCSD(T)/$150$au  & -542  \\
  81e-CCSD(T)/$10^4$au & -558  \\ 
  \hline\hline
 \end{tabular}
\end{table}

To check the convergence with respect to electron correlation effects we performed a series of 
successive 21-electron coupled cluster calculations within the TZ basis set (see Table \ref{TABcc2}). 
In these calculations two sets of atomic bispinors were used. The first one was obtained within the 
DCHF approximation where the Fock operator is defined by averaging over $6p_{j=1/2}^1$ and $6p_{j=3/2}^1$ open-shell electronic configurations as  in the CI case above.
The second one was obtained within the closed-shell DCHF
method for the Tl$^+$ cation. One can 
see that CC values gives almost identical result for each set at any level. Besides, the contribution of 
correlation effects beyond the CCSD(T) model
is almost negligible in the considered case. We considered models up to coupled cluster with Single, Double, Triple and perturbative Quadruple cluster amplitudes, CCSDT(Q).

Contribution of the effect of the Breit interaction on R has been estimated in reference \cite{porsev_safronova_kozlov_tl2012} as 0.36\%. Based on the uncertainties discussed above we conservatively estimate the uncertainty of our final CC value for $R$ to be 
less than 5\%.

\begin{table}
\caption{Values of $R$ calculated at different level of theory with correlation of 21 electrons of Tl,
         cvTZ basis set. 
         Calculations were performed using DCHF spinors for the neutral Tl atom ($V^N$ potential) and 
         for the Tl$^+$ cation ($V^{N-1}$ potential) cases.
In both cases the external field perturbs both the spinor coefficients and the CC amplitudes.         
         }
\label{TABcc2} 
\begin{tabular}{lll}
\hline\hline
         & $V^N$ & $V^{N-1}$    \\  \hline         
DCHF      
& -418  & -402 \\
CCSD     & -531  & -530 \\
CCSD(T)  & -521  & -522 \\
CCSDT    & -523  & -523 \\
CCSDT(Q) & -522  & -522 \\
\hline\hline
\end{tabular}
\end{table}

\subsection{Discussion in Comparison with Literature Results}

Our present best results are shown in Table \ref{TAB3} in comparison with previous work.
The earlier controversy between different groups over results for $R$(Tl) can be condensed into three main
points which we address one by one.

\begin{table}
 \caption{Comparison with literature values \label{TAB3} }
 \begin{tabular}{l|ccc}
 Work   & $\alpha_{d_e}$ & $\alpha_{C_S} [10^{-18} e$ cm$]$ & $A_{||}$({$^{205}$Tl}) [MHz]  \\ \hline\hline
  \multicolumn{3}{c}{Literature values} \\ \hline
  Khriplovich {\it{et al.}} \cite{khriplovich_lamoreaux}       &  $    $  & $5.1$   &         \\
  Johnson {\it{et al.}} \cite{johnson_tlEDM_1986} Norcross potential &  $-562$  & $   $   & $-18764$ \\
  M{\aa}rtensson-Pendrill {\it{et al.}} \cite{mar-pendrill-lindroth_EPL1991} 
                                                     &  $-600 \pm 200$  & $7 \pm 2$ &         \\ 
  Liu {\it{et al.}} \cite{Liu_Kelly_Tl_enhancement}            &  $-585$  &         &         \\
  Dzuba {\it{et al.}} \cite{dzuba_flambaum_Cs_Tl_2009}         &  $-582$  &$7.0\pm 0.2$ & $21067$ \\
  Nataraj {\it{et al.}} (CCSD(T)) \cite{Nat_PRL_EDM_Tl}        &  $-470$  &         & $21053$ \\
  Sahoo {\it{et al.}} (CCSD(T)) \cite{Sahoo_CsTl_2008}         &          &  $4.06$ & $21026$ \\
  Porsev {\it{et al.}} \cite{porsev_safronova_kozlov_tl2012}   &  $-573$  &         & $22041$ \\ \hline
  This work CI                                                 &  $-539$  &  $6.61$ & $20614$ \\
  This work CC                                                 &  $-559$  &  $6.77$ & $21172$ \\ \hline
  Experiment \cite{Tl_hyperfine_exp1998}                       &          &         & $21310.8 \pm 0.0$
 \end{tabular}
\end{table}

\subsubsection{Basis sets}

From the results in Tables \ref{TAB2} and \ref{TABcc1} it is evident that a large atomic basis set,
at least of quadruple-zeta quality, must be used for obtaining very accurate interaction constants.
The results in \ref{TAB1} and \ref{TABcc1} obtained with our correlation methods demonstrate that the 
basis set used by Nataraj \etal\ in ref. \cite{Nat_PRL_EDM_Tl} fulfills this requirement, yielding
interaction constants that are very close to those obtained with Dyall's cvQZ basis set and the
same correlation expansion. The earlier 
suggestion
%
of Porsev \etal\ about an inadequate basis set
used in ref. \cite{Nat_PRL_EDM_Tl} can, therefore, be excluded as a possible reason for the outlier
result in ref. \cite{Nat_PRL_EDM_Tl}.

\begin{table}
 \caption{R, S, and A for Tl atom \label{TAB1} }
 \begin{tabular}{l|ccc}
 Model/virtual cutoff                         & $R$     & $S$ [a.u.] & $A_{||}$({$^{205}$Tl}) [MHz]    \\ \hline\hline
  \multicolumn{4}{c}{Nataraj universal} \\ \hline
  CAS3in4                                      & $-483$  & $339$ & $18800$ \\
  CAS3in4\_SD3/6au                             & $-597$  & $419$ & $19035$ \\
  CAS3in4\_SD3/20au                            & $-523$  & $367$ & $19174$ \\
  CAS3in4\_SD3/45au                            & $-595$  & $419$ & $19060$ \\
  CAS3in4\_SD3/Nat100                          & $-595$  & $418$ & $19060$ \\
  CAS3in4\_SDT3/45au                           & $-596$  & $419$ & $19069$ \\
  CAS3in4\_SD3/130au                           & $-595$  & $418$ & $19060$ \\
  CAS3in4\_SD3/200au                           & $-595$  & $418$ & $19060$ \\
  SD18\_CAS3in4\_SD21/45au                     & $-510$  & $361$ & $19864$ \\ \hline\hline 
 \end{tabular}
\end{table}

\subsubsection{Treatment of correlation effects by the many-body method}
It is claimed in reference \cite{Nat_PRL_EDM_Tl} that the treatment of electron correlation effects were
more complete than in references \cite{Liu_Kelly_Tl_enhancement} and \cite{dzuba_flambaum_Cs_Tl_2009}.
We have therefore first attempted to reproduce the electron EDM enhancement calculated by Nataraj \etal\
by using the same many-body Hamiltonian and EDM operator, the same atomic basis set (``Nataraj universal'') 
and the
the same method, CCSD(T).
A persisting difference with the approach of Nataraj \etal\ 
is the use of CC amplitudes for the closed shells of neutral
Tl (our case) or the closed shells of the singly-ionized Tl$^+$ (Nataraj case). 
These results are shown in Table \ref{TABcc1} under the label ``$V^{N-1}$''.
Our calculation of the hyperfine constant $A_{||} = 21071$ [MHz] reproduces the value of Nataraj \etal\
which is $A_{||} = 21053$ [MHz] almost precisely (residual difference of less than $0.1$\%). 
However, using the same wavefunction we obtain $R = -550$ which differs from the value of Nataraj \etal\ 
by $17$\%. Our CC result for $R$ is in accord with similar calculations using the large cvQZ basis set,
in accord with the present best CI result ($R=-539$ which after correction for core correlations from the
innermost $28$ electrons,
according to the results in Table \ref{TABcc3},
becomes $R=-555$) and in good agreement with the best results of Liu \etal\ 
\cite{Liu_Kelly_Tl_enhancement}, Dzuba \etal\ \cite{dzuba_flambaum_Cs_Tl_2009}, and Porsev \etal\ 
\cite{porsev_safronova_kozlov_tl2012}, see Table \ref{TAB3}. 
The correct evaluation of the electron
EDM enhancement in our codes has been assured by comparative tests of the independent implementations
of present CI and CC, as well as with the \verb+DIRRCI+ module \cite{luuk_dirrci,fleig_nayak_eEDM2013} in 
the DIRAC program package.
All three independent implementations produce the same values of $R$ for small test cases using
Full CI / Full CC expansions. These findings strongly suggest that the CC wavefunctions used by us and
by Nataraj \etal\ are almost identical, but that the evaluation of $R$ in reference \cite{Nat_PRL_EDM_Tl}
is flawed.

Since correlation effects have been treated at a 
very similar (but physically more accurate) level in the present work as in ref. \cite{Nat_PRL_EDM_Tl} and the result is very different,
the claim of correlation effects being responsible for the large difference between previous results is untenable.
%

\subsubsection{Use of $V^{N}$, $V^{N-1}$, and $V^{N-3}$ potentials}

First, given a fixed atomic basis set and a fixed many-body Hamiltonian \footnote{In the present and previous
works the Dirac-Coulomb picture is employed where negative-energy states are implicitly or explicitly excluded
from the orbital/spinor space which is used as a basis for the many-body expansion.}, the Full CI 
expansion delivers the exact solution in the $N$-particle sector of Fock space \cite{helgaker-book}, {\it{independent
of the orbital/spinor basis used for this Full CI expansion}}. This implies that a many-body expansion that
closely approximates the Full CI expansion, such as CCSDT or CCSDT(Q), must also be nearly independent of
the employed Dirac-Fock potential. 

Our results in Table \ref{TABcc2} clearly confirm this conjecture and demonstrate that even in the more
approximate CCSD expansion the electron EDM enhancement factor $R$ is almost independent ($0.2$\% difference)
of the underlying
spinor set. As the many-body expansion becomes more approximate, such as in the CI model SD18\_CAS3in4\_SD21
(see Table \ref{TAB2}) basic theory leads us to expect that the difference in $R$ should increase which is
indeed the case (roughly $2$\% difference). Adding external Triple excitations to the CI expansion, model
SD18\_CAS3in4\_SDT21, quenches the difference to a mere $1.2$\%, again in accord with expectation. 
Even the use of a $V^{N-3}$ potential (\ie\ spinors optimized for the Tl$^{3+}$ system) changes $R$ by
only about $3$\% relative to spinors for the neutral atom in the SD18\_CAS3in4\_SDT21 model. This difference
is expected to be even smaller in CC models.

Despite of the unimportance of the employed spinor set in highly-correlated calculations, we have used
the physically most accurate spinors for the neutral Tl atom in obtaining our best final results. The ratios of our calculated $\cal{P,T}$-odd interaction
constants are $\left|\frac{\alpha_{d_e}}{\alpha_{C_S}}\right|$(CI) $= 81.5 \frac{1}{10^{-18} [e {\text{cm}}]}$ and $\left| \frac{\alpha_{d_e}}{\alpha_{C_S}} \right|$(CC) $= 82.6 \frac{1}{10^{-18} [e {\text{cm}}]}$ which agree well with the
analytical value of Dzuba \etal\ \cite{PhysRevA.84.052108,PhysRevA.85.029901}
of $\left| \frac{\alpha_{d_e}}{\alpha_{C_S}} \right|$(an.) $= 89 \frac{1}{10^{-18} [e {\text{cm}}]}$.

\section{Conclusions}
\label{SEC:CONCL}

We conclude from our findings that the result of Nataraj \etal\ in ref. \cite{Nat_PRL_EDM_Tl}
is unreliable and should be excluded from the dataset used to constrain the ${\cal{CP}}$-odd
parameters $d_e$ and $C_S$. Likewise, the result by Sahoo \etal\ \cite{Sahoo_CsTl_2008} (see Table \ref{TAB3})
-- presumably obtained with a similar code as R(Tl) by Nataraj \etal\ -- is also significantly
too small and should be excluded from the reliable dataset. A further proof of this conclusion
is the ratio obtained using the values from refs. \cite{Nat_PRL_EDM_Tl} and \cite{Sahoo_CsTl_2008} which amounts
to $\left| \frac{\alpha_{d_e}}{\alpha_{C_S}} \right|$(CC Nataraj/Sahoo) $= 115.8 \frac{1}{10^{-18} [e {\text{cm}}]}$
which deviates from the analytical ratio \cite{PhysRevA.84.052108,PhysRevA.85.029901} by $30$\%.

\vspace*{0.5cm}
\noindent
\begin{minipage}{16.2cm}
 \begin{minipage}{7.0cm}
  \begin{center}
    \includegraphics[angle=0,width=6.2cm]{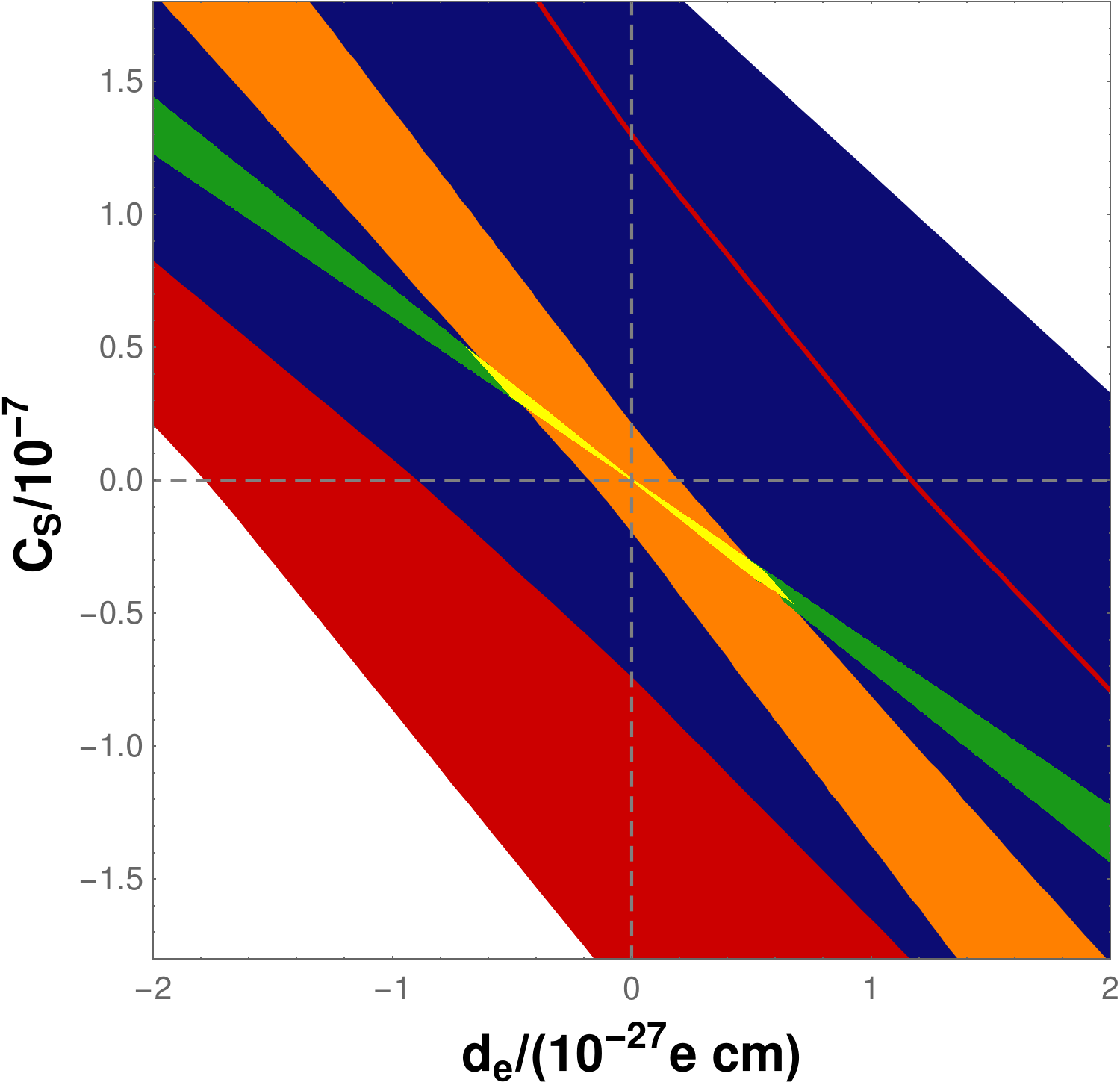}
  \end{center}
 \end{minipage}
  \hfill
 \begin{minipage}{8.8cm}
  The panel is the updated version of the one shown in the introduction, using the
  dataset of reliable calculations of $\alpha_{d_e}$ and ${\alpha_{C_S}}$  
  for the Tl atom. The strongly reduced uncertainty of atomic interaction constants for Tl
  leads to a discernable shrinking of the associated parameter surface (blue), but does not
  lead to modified constraints. The essential reason for this is the extremely high
  sensitivity of the experiments on ThO (green) and HfF$^+$ (orange).
  However, tighter constraints on $d_e$ and $C_S$ can be obtained by including experimental and theoretical
  results for closed-shell atomic systems as discussed in ref. \cite{FleigJung_JHEP2018}.
\end{minipage}
\end{minipage}


\section{Acknowledgements}
We thank Martin Jung (Torino) for providing updated plots and for helpful discussions.
Huliyar Nataraj is thanked for sharing many technical details of his calculations with us.
Electronic structure calculations were partially carried out using resources of the collective usage centre Modeling and predicting properties of materials at NRC ``Kurchatov Institute'' -- PNPI. L.S. was supported by the Russian Science Foundation Grant No. 19-72-10019.


\newcommand{\Aa}[0]{Aa}

\end{document}